\documentclass[reprint,nofootinbib,nobibnotes,amsmath,amssymb,aps,prresearch,floatfix,twocolumn,longbibliography]{revtex4-1}
\usepackage{CJK}
\usepackage{latexsym}
\usepackage{hyperref}
\usepackage{times}
\usepackage{graphicx}
\usepackage{amsmath}
\usepackage{color,subfigure}

\newcommand{\disp}[1]{Eq. (\ref{#1})}

\newcommand{\refdisp}[1]{Ref. [\onlinecite{#1}]}
\newcommand{\figdisp}[1]{Fig. \ref{#1}}

\newcommand{\beq}{\begin{eqnarray}}
\newcommand{\eeq}{\end{eqnarray}}

\usepackage{dcolumn}
\usepackage{bm}
\usepackage{ textcomp }
\usepackage{color}
\usepackage{amssymb}
\usepackage{xcolor}
\usepackage{easyReview}
\usepackage{mathdots}
\usepackage{float}
\usepackage{lineno}
\usepackage{todonotes}
\usepackage{array}
\usepackage{xcolor,colortbl}
\begin{document}
\begin{CJK*}{GB}{} 

\title{Topological Mott Insulator at Quarter Filling in the Interacting Haldane Model}
\author{ Peizhi Mai$^{1}$, Benjamin E. Feldman$^{2,3,4}$, Philip W. Phillips$^{1,\dagger}$ }

\affiliation{$^1$Department of Physics and Institute of Condensed Matter Theory, University of Illinois Urbana-Champaign, Urbana, IL 61801, USA}
\affiliation{$^3$Geballe Laboratory of Advanced Materials, Stanford, CA 94305, USA}
\affiliation{$^4$Department of Physics, Stanford University, Stanford, CA 94305, USA}
\affiliation{$^5$Stanford Institute for Materials and Energy Sciences, SLAC National Accelerator Laboratory, Menlo Park, CA 94025, USA}


\begin{abstract}
 \textbf{While the recent advances in topology have led to a classification scheme for electronic bands described by the standard theory of metals, a similar scheme has not emerged for strongly correlated systems such as Mott insulators in which a partially filled band carries no current. By including interactions in the topologically non-trivial Haldane model, we show that a quarter-filled state emerges with a non-zero Chern number provided the interactions are sufficiently large.  We first motivate this result on physical grounds and then by two methods: analytically by solving exactly a model in which interactions are local in momentum space and then numerically through the corresponding Hubbard model. All methods yield the same result: For sufficiently large interaction strengths, the quarter-filled Haldane model is a ferromagnetic topological Mott insulator with a Chern number of unity.  Possible experimental realizations in cold-atom and solid state systems are discussed.}
\end{abstract}
\maketitle
\end{CJK*}

\section{Introduction}

To a large extent, topology and strong electron correlation, the two pillars of quantum materials, have evolved essentially independently. The highly successful  classification scheme\cite{haldane,kanemele} of band insulators into topologically trivial or non-trivial assumes the electrons are non-interacting and hence  only the details of the band structure are relevant.  In the other extreme lies the physics of doped Mott insulators, the parent state of copper-oxide high-temperature superconductors.  As Mott insulation traditionally obtains in a half-filled band, it naturally implies a breakdown of the single-particle concept. Further, Mott insulation is generally formulated in real-space from on-site interactions and topological invariants require a momentum-space picture.  How is it then possible to formulate a topological classification scheme for Mott insulators?  

\begin{figure}
    \centering
    \includegraphics[width=0.4\textwidth]{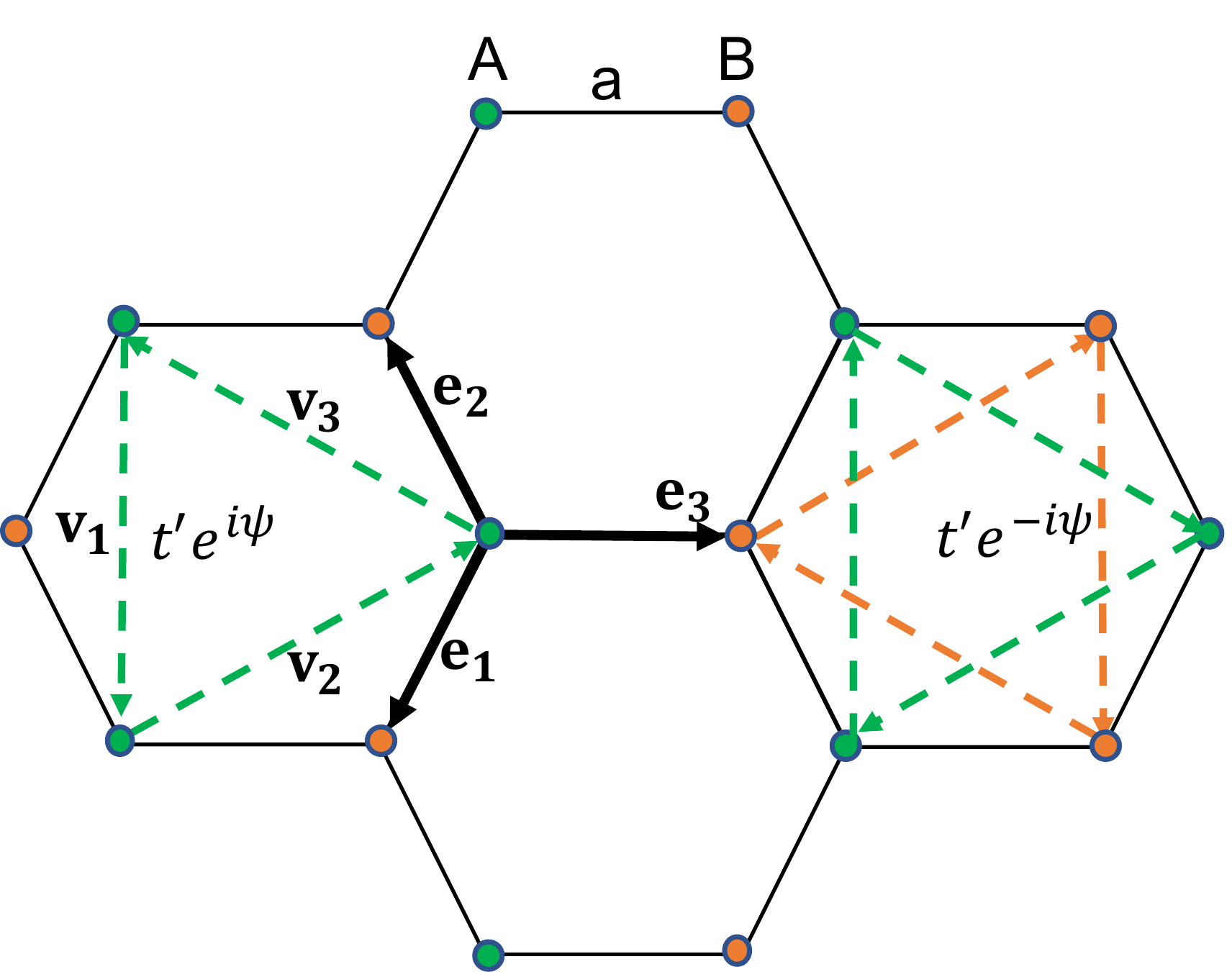}
    \caption{Hopping vectors and parameters for the Haldane Hamiltonian.  Because of the complexified hopping, $t'$ in \disp{HaldaneH2}, there is a net magnetic flux in each triangle which cancels out when averaged over the hexagon. 
    }
    \label{fig:haldane}
\end{figure}

To uncloak topological features of Mott physics, we note that although traditionally topology is concerned with the equivalent ways of distorting a geometric object, it is the algebraic formulation that is most relevant to condensed matter systems.  Consider the energy spectrum of some quantum mechanical system. Any continuous deformation of the underlying Hamiltonian, keeping the same boundary conditions, which leaves the bulk physics unchanged is inherently topological. It is this realization\cite{haldane,kanemele} that accounts for the stability of edge currents in the quantum Hall\cite{L0} effect and protects the surface states\cite{kanemele,fukanemele,jmoore,spinhallscz,rahulroy1,bhz} in topological insulators from acquiring a gap.  While it is a large perpendicular magnetic field that creates the edge currents in quantum Hall systems, the same physics obtains even in the absence of a net magnetic field. In practise, topologically non-trivial insulation arises either from the relativistic effect resulting from the  a spin-dependent force or spin-orbit coupling\cite{kanemele,fukanemele,jmoore,spinhallscz,rahulroy1,bhz}  or  complex hoppings on a honeycomb\cite{haldane} as in \figdisp{fig:haldane} or a square lattice\cite{bhz}.  The hallmark of bands with non-trivial topology is a non-zero Chern number\cite{thouless} which is computable entirely from the single-particle states.

Of course there are instances in which topology and interactions merge as in the fractional\cite{laughlin,haldanesym} quantum Hall effect (FQHE).  Here, no single-particle description can encode the fractional charge or statistics of the quasiparticles in the fractional regime of the QHE.  However, while it is the interactions that mediate the FQHE, the gap is set by the magnetic field rather than the interactions as in traditional Mott insulators.  Recently, numerous examples have arisen of fractional Chern insulators in which features of the FQHE arise in the absence of Landau levels\cite{chamonprl,neupert,parameswaran}. In such systems, gaps\cite{chamonprl,neupert,parameswaran} open at fractional fillings and are set by the interaction strength. Such gaps seem to only arise in the flat-band limit in which the $\Delta\gg U (\text{or}~V)\gg W$, where $\Delta$ is the non-interacting topological gap, $U$ the interaction strength and $W$ the bandwidth. Outside the flat-band limit, the study on the interacting spinless system\cite{Grushin,Zhu} reveals a fractional Chern insulator driven by a strong nearest-neighbor interaction. Hence, a key question arises: outside the flat-band limit, are there general cases of Mottness-driven topological phases in a spinful system?  Such cases would provide an analogue of a Mott insulator driven by the strong correlations but with topological signatures, for example a non-zero Chern number.  



To achieve our goal of finding a topological Mott insulator, we put interactions into the spinful Haldane model\cite{haldane} of Chern insulators.  In so doing, we uncover an overlooked phase at $1/4$-filling which is a Mott insulator with a gap set by the non-trivial topology of the Haldane model. Prior work revealing inter-relations between Mott physics and topology employed perturbative methods on models with spin-orbit and Hubbard-type interactions\cite{pesinbalents, Herbut1, Herbut2}, numerical methods on the spinless and spinful\cite{Vanhala,Shao,Imriska,Mertz} Haldane-Hubbard model, models with band-touching in 3-dimensions\cite{herbut} and even bosonic systems\cite{kuno}.  Our work here is focused on making exact statements regardless of filling.  Just as topological insulation can be thought of as a completion of band structure by the inclusion of spin-orbit coupling, our work here on topological Mott insulation away from half-filling is a natural outgrowth of putting interactions into the spinful Haldane model. Recall the spinful Haldane model has two spin-degenerate bands separated by a topological gap.  Hence, at half-filling the system is an insulator.  Now consider including interactions specifically repulsive interactions between spin-up and -down electrons.  The singly and doubly occupied states in each band are no longer degenerate.  For sufficiently large interactions, the doubly occupied bands will be pushed up to high energy whereas the singly occupied states in both the lower and upper Haldane bands will remain fixed in energy.  The gap between them is wholly set by topology.  Each of these bands now just has half the number of electrons relative to the non-interacting case.  At 1/4-filling, the chemical potential lies in the gap between the two singly occupied bands resulting in an insulator.  Since it is the interactions that pushed half of the states to high energy, the resultant state is a topological Mott insulator.

We prove this physics by considering two models, one local in momentum space as in the Hatsugai-Kohomoto (HK) model\cite{HK} and the other local in position, the Hubbard interaction.   While these interactions are extremes of one another, they both share a key ingredient.   Namely, they both break the local-in-momentum space $\mathbb Z_2$ symmetry of a Fermi surface. That is, electrons on a Fermi surface are invariant\cite{AHO4,HKnp2} to an interchange of particles  and holes for a single spin species: $c_{p\uparrow}\rightarrow c_{p\uparrow}^\dagger$. Here, $c_{\bf p\sigma}^\dagger$ ($c_{\bf p\sigma}$) creates (annihilates) an electron with momentum $\bf p$ and spin $\sigma$.  Interactions, in general, do not preserve this symmetry and hence the breaking of the $\mathbb Z_2$ symmetry is a tell-tale sign that interactions are the root cause of the well known phenomenon of particle-hole asymmetry\cite{davis2004,lanzarapha,ong,singhpha,asym} in the particle-addition and removal spectrum widely observed in correlated electron systems.  To further buttress our previous arguments\cite{HKnp2} that breaking this symmetry establishes a fixed point and hence the nature of the interactions which accomplishes this is irrelevant, we show that essentially the same results obtain for the Hubbard as well as the HK models.  We show explicitly that interactions induce a topologically non-trivial Mott insulating phase at quarter filling in the Haldane model for both the HK and Hubbard models that can be understood with the simple picture presented in the previous paragraph.  The advantage of the HK model is that all the calculations can be performed analytically.  Our result then constitutes the first example of an exact demonstration of a topological phase driven by interactions.  From the DQMC calculations with  the Hubbard interaction, we are able to show that ferromagnetic correlation obtains in the quarter- and three-quarter-filled Haldane-Hubbard model.  

\section{Exactly solvable model for interacting Chern insulators}

\begin{figure*}[ht]
    \centering
    \includegraphics[width=\textwidth]{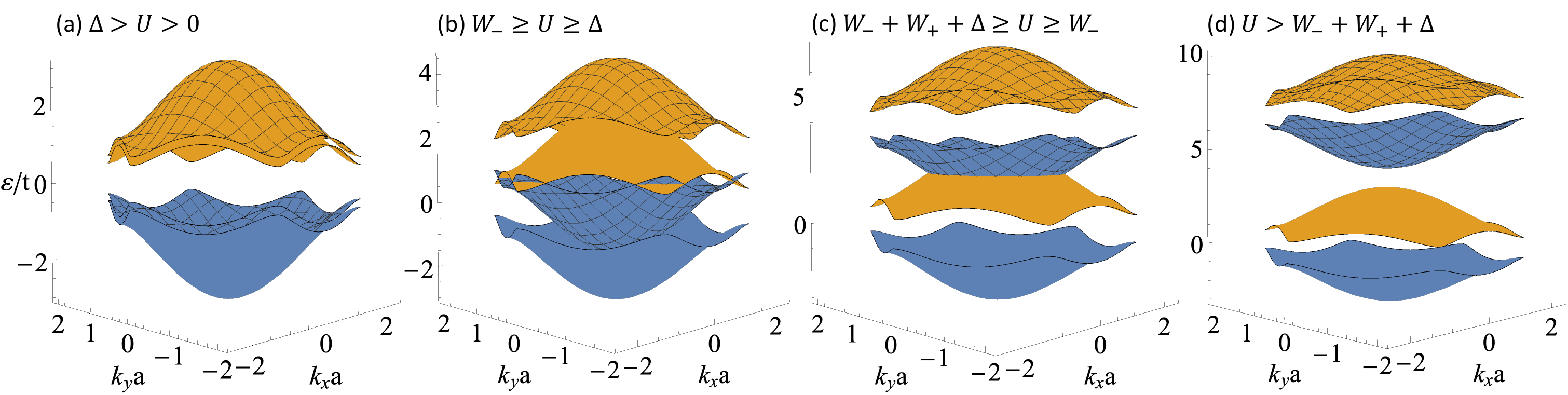}
    \caption{Band structure for the Haldane-HK model in \disp{HHK} with $t'/t=0.1, \psi=-\pi/2, M=0$ at zero temperature. $\Delta\approx1.04t$ is the non-interacting topological gap. $W_{+(-)}\approx2.48t$ correspond to the bandwidth of the Haldane upper (lower) band. Different phases evolve as the interaction strength increases:  (a) $1/2$-filled topological insulator for $\Delta>U>0$ ($U/t=0.2$), (b) a metal for $W_-\geq U \geq \Delta$ ($U/t=1.5$), (c) $1/4$-filled Mott insulator with non-zero Chern number for $W_-+W_++\Delta\geq U\geq W_-$ ($U/t=4$), and (d) $1/4$-filled insulator for $U>W_-+W_++\Delta$ ($U/t=7$). The blue (or orange)  color represents Chern number $C=-1 ($or $ 1)$. The meshed band consists of only doubly occupied states, while the unmeshed band is singly occupied.
    }
    \label{fig:QHEphase}
\end{figure*}

 For spinful electrons, the Haldane model\cite{haldane} with the hoppings designated in \figdisp{fig:haldane} takes a simple form
\beq
H=\sum_{{\bf k},\sigma}\big[(\varepsilon_{+,{\bf k}}-\mu)n_{+,{\bf k}\sigma}+(\varepsilon_{-,{\bf k}}-\mu)n_{-,{\bf k}\sigma}\big],\label{HaldaneH1}
\eeq
in momentum space. Here $n_{\pm,{\bf k}\sigma}$ is the number of electrons in the Haldane upper ($+$) and lower ($-$) bands respectively, $\mu$ is the chemical potential and
\beq
\varepsilon_{\pm,{\bf k}}=h_0({\bf k})\pm\sqrt{h_x^2({\bf k})+h_y^2({\bf k})+h_z^2({\bf k})}
\eeq
with
\beq
\begin{aligned}
&h_0({\bf k})=-2t'\cos{\psi}[\sum_{i=1}^3 \cos({\bf k}\cdot {\bf v}_i)],
\\&h_x({\bf k})=-t[\sum_{i=1}^3 \cos({\bf k}\cdot {\bf e}_i)],
\\&h_y({\bf k})=-t[\sum_{i=1}^3 \sin({\bf k}\cdot {\bf e}_i)],
\\&h_z({\bf k})=M-2t'\sin{\psi}[\sum_{i=1}^3 \sin({\bf k}\cdot {\bf v}_i)],\label{HaldaneH2}
\end{aligned}
\eeq
represent the Haldane upper and lower bands. As depicted in \figdisp{fig:haldane}, $t$ and $t'$ stand for the first and second neighbor hopping respectively, ${\bf e}_i$ and ${\bf v}_i$ are the bonding vectors for nearest neighbors and next nearest neighbors respectively, $M$ is the Semenoff\cite{semenoff} mass. The hopping parameters are shown in \figdisp{fig:haldane}. As is well known\cite{haldane}, the half-filled system is a topological insulator with Chern number $C=\pm 2$ (due to spin degeneracy) when $\big|M\big|<3\sqrt{3}\big| t'\sin{\psi}\big|$ or a topologically trivial insulator with Chern number $C=0$ when $\big|M\big|>3\sqrt{3}\big| t'\sin{\psi}\big|$.  Consequently, the gap between the degenerate spinful Haldane upper and lower bands is set by $\Delta=\min(6\sqrt{3}\big|t'\sin{\psi}\big|,2)$ for small $t'$.  The existence of non-trivial topology at half-filling is also shared with the model studied by Pesin and Balents\cite{pesinbalents}.

We now introduce interactions in the spirit of Hatsugai-Kohmoto (HK) interaction\cite{HK,HKnp1,HKnp2} leading to the Hamiltonian,
\beq
\begin{aligned}
H&=\sum_{{\bf k},\sigma}\big[(\varepsilon_{+,{\bf k}}-\mu)n_{+,{\bf k}\sigma}+(\varepsilon_{-,{\bf k}}-\mu)n_{-,{\bf k}\sigma}\big]
\\&+U\sum_{\bf k}(n_{+,{\bf k}\uparrow}n_{+,{\bf k}\downarrow}+n_{-,{\bf k}\uparrow}n_{-,{\bf k}\downarrow}), \label{HHK}
\end{aligned}
\eeq
describing interacting electrons in the Haldane model.  The primary role played by the interaction is to lift the $\mathbb Z_2$ symmetry of the non-interacting Fermi surfaces.  What this effectively does is introduce Mottness in the form of singly occupied states below the chemical potential. It is a simplicity of the HK model that such strong correlation is introduced without mixing the original non-interacting eigenstates.  As a result, momentum $\bf k$ is still a good quantum number and the Green function can be written down immediately\cite{HKnp1} as
\beq
\begin{aligned}
G_{\pm,{\bf k}\sigma}(i\omega_n)&\equiv	-\int_0^\beta d\tau\langle c_{\pm,{\bf k}\sigma} (\tau) c^\dagger_{\pm,{\bf k}\sigma}(0) \rangle \text{e}^{i\omega_n\tau} \\
G_{\pm,{\bf k}\sigma}(i\omega_n\rightarrow \omega)&=\frac{1-\langle n_{\pm,{\bf k}\bar{\sigma}}\rangle}{\omega+\mu-\varepsilon_{\pm,{\bf k}}} + \frac{\langle n_{\pm,{\bf k}\bar{\sigma}}\rangle}{\omega+\mu-(\varepsilon_{\pm,{\bf k}}+U)}.\label{GF}
\end{aligned}
\eeq
From the Green function, we see immediately the effect of the correlations.  Each of the lower, $\varepsilon_{-,{\bf k}}$ and upper bands, $\varepsilon_{+,{\bf k}}$ will now be split into a singly and doubly occupied subband. Since, there is already a topological gap at half-filling, in the presence of interactions as the doubly occupied bands move up in energy, the singly-occupied bands can never get closer than  $\Delta$.  Hence, at quarter-filling, a gap obtains and the non-trivial topology of the Haldane bands persists.  It is the emergence of $1/4$-filled Mott insulating states in a system that has a topologically engineered gap at  half-filling that is the principal conclusion of this paper.  To see how this state of affairs plays out,  we plot the band structure in
 \disp{GF}.  As an example, we set $t'/t=0.1$, $\psi=-\pi/2$, $M=0$ without loss of generality so that the Haldane lower (upper) band has a Chern number $C=-1 (+1)$ for each spin and hence the non-interacting half-filled system is a topological insulator with Chern number $C=-2$. We use $W_{+(-)}$ for the bandwidth of the Haldane upper (lower) band and $\Delta$ for the Haldane gap. For this parameter set, $W_{+}=W_{-}>\Delta$. Turning on $U$ now results in four different phases for different interaction strengths,  all of which are shown in \figdisp{fig:QHEphase}. The blue and orange bands carry Chern $C=-1$ and $C=1$, respectively. The interaction $U$ only affects bands with the same color, making the blue and orange bands split into a singly occupied lower subband (unmeshed) and doubly occupied upper subband (meshed). However, the energy separation between the two unmeshed orange and blue bands remains fixed. This leads to a non-trivial interplay between topology and interaction strength.   In \figdisp{fig:QHEphase}(a) when $U$ is small, the non-interacting band structure is mostly intact, and the system is a topological insulator at half-filling with Chern number $C=-2$. Increasing $U$ to $W_{-}\geq U \geq \Delta$ as shown in \figdisp{fig:QHEphase}(b) leads to a conductor with no gap for any density other than in the empty or full band limits.  Increasing $U$ further to $W_{-}+W_{-}+\Delta \geq U > W_{-}$ as shown in \figdisp{fig:QHEphase}(c), so that the two unmeshed singly occupied bands have no overlap with their doubly occupied same-color meshed-partners, results in a gap opening at $1/4$- and $3/4$-filling. While it is interactions that lead to a splitting of the bands, it is topology that dictates the separation between the two unlike-colored bands (meshed or otherwise).  Consequently, the gap at $1/4$- and $3/4$-filling is set by the Haldane gap, $\Delta$.   The Chern number in either case is $C=-1$ and hence, this strongly correlated non-metallic state is actually a topological Mott insulator. Note at quarter-filling, the system consists of only singly occupied states as a result of strong correlation.   We will resort to a Hubbard interaction to settle the nature of the spin correlations.  Now consider increasing $U$ further so that $U>W_{-}+W_{-}+\Delta$.  While the situation for $1/4$- and $3/4$-filling remains qualitatively unchanged, a gap opens at half-filling and the system at this density becomes a topologically trivial Mott insulator with only singly occupied states and Chern number $C=0$. Although the features might change quantitatively in \figdisp{fig:QHEphase}(b) and (c) for intermediate values of $U$ with a different parameter set $(t',\psi,M)$, \figdisp{fig:QHEphase}(d) is generally true for all parameters.  This means when the correlation is strong enough, the system becomes an insulator at $1/4$- and $3/4$-filling with Chern number $C=C_0/2$ where $C_0$ is the Chern number at half-filling without interaction, and a topologically trivial insulator at half-filling with $C=0$.  Consequently, from the Hamiltonian \disp{HHK}, we are able to analytically describe the transition from a metal to a quarter-filled topological Mott insulator  should $U$ be the dominant energy scale.  This constitutes the first analytically solvable model for the onset of topological phases in the presence of interactions. While \disp{HHK} is in the parameters ($t',\phi,M$)-dependent Haldane orbital basis, it is also possible to write down an analytically solvable Hamiltonian in sub-lattice basis that contains essentially the same physics for strong interaction (see supplement). Due to its simplicity, we refer \disp{HHK} as the analytically solvable Hamiltonian for interacting Haldane electrons. Note in the previous work, topologically induced interacting phases emerged only in the flat-band limit\cite{neupert,parameswaran}.  \figdisp{fig:QHEphase} illustrates that no such constraint is necessary for the $1/4$ or $3/4$-filled topological insulating phases.  As a final note, it is worth pointing out that topological states in the $1/4$- and $3/4$-filling in the Kitaev-Hubbard models\cite{senechal} exist already in the non-interacting system and are impervious to interactions.    What is new here is that the interactions induce a topological Mott insulator at $1/4$- and $3/4$-filling without affecting any details of the topology.

\section{Numerical study on Haldane-Hofstadter-Hubbard model}

It is natural to ask whether these conclusions persist in the more traditional Haldane-Hubbard model in which the interactions result in the mixing of the momentum states. Previous numerical studies \cite{Vanhala,Shao,Imriska,Mertz,Varney1,Varney2} on the Haldane-Hubbard model focused mostly on the half-filling cases and showed a transition from a  Chern insulator to a topologically trivial Mott insulator. However, the region away from half-filling has rarely been explored. The deep question arises: do the $1/4$- and $3/4$-filled insulating states still appear when the interaction term mixes the orbitals that comprise the Haldane topological bands?  The interaction used previously preserves the Haldane basis and hence, all it  can do is to move the doubly occupied states up, keeping the distance between the singly occupied (unmeshed) sectors the same.  What we want to show here is that the same essential conclusion is true for the Hubbard model.  

The Hamiltonian for the Haldane-Hofstadter-Hubbard model is
\beq
\begin{aligned}
\label{Eq:HubbardBfield}
    H&= -\sum_{{\bf i}{\bf j}\sigma} t_{{\bf i},{\bf j}}\exp(i \phi_{{\bf i},{\bf j}} ) c^\dagger_{{\bf i}\sigma}c^{\phantom\dagger}_{{\bf j}\sigma} 
    -\mu\sum_{{\bf i},\sigma} n_{{\bf i}\sigma}\\
    &+ M\sum_{{\bf i}_A,{\bf i}_B,\sigma} (n_{{\bf i}_A\sigma}- n_{{\bf i}_B\sigma}) + U\sum_{{\bf i}}(n_{{\bf i}\uparrow}-\frac{1}{2})(n_{{\bf i}\downarrow}-\frac{1}{2}),
\end{aligned}
\eeq
where $t_{{\bf i},{\bf j}}$ represents the nearest neighbor hopping $t$ (set to $1$ as the energy scale) and next nearest neighbor hopping $t'\text{e}^{\pm i\psi}$, ${\bf i}_{A(B)}$ means the lattice sites in the A (or B) sublattice, as shown in \figdisp{fig:haldane}. Throughout this study, we set the Haldane phase $\psi=-\pi/2$ without loss of generality. Due to the presence of an external uniform magnetic field, we use the Peierls substitution\cite{peierls} to introduce the phase through the flux threading,
\beq
\phi_{{\bf i},{\bf j}}=\frac{2\pi }{\Phi_0} \int_{r_{\bf i}}^{r_{\bf j}} {\bf A}\cdot d{\bf l},
\eeq
where $\phi_0=h/e$ represents the magnetic flux quantum and the integration is over the straight line path from site ${\bf i}$ to ${\bf j}$. Here we apply an external magnetic field to measure the Chern number of the incompressible states at high temperature.  As we will see, all the notable features are present (and stronger) for the smallest fluxes, thereby minimizing the effect of the external field which computationally reduces the finite size effects\cite{Mai} at low temperature.

We use the DQMC method\cite{Blankenbecler,Hirsch,White} to simulate this model on an $N_{\text{site}}=6\times6\times2$ cluster. The factor of $2$ accounts for the two sublattices in the honeycomb lattice. The modified periodic boundary conditions in \refdisp{Assaad} are adjusted for the honeycomb lattice\cite{Mai}. The flux quantization condition $\Phi/\Phi_0=n_f/N_c$ (with $n_f$ an integer and $N_c$ the number of unit cells) needs to be fulfilled for a single-value wave function. We choose the symmetric gauge ${\bf A}=(x\hat{y}-y\hat{x})B/2$ for this calculation. Due to the Fermionic sign problem, we focus on the inverse temperature $\beta=3/t$ and vary other parameters $t', M, U$. 

\begin{figure}[ht!]
    \centering
    \includegraphics[width=0.5\textwidth]{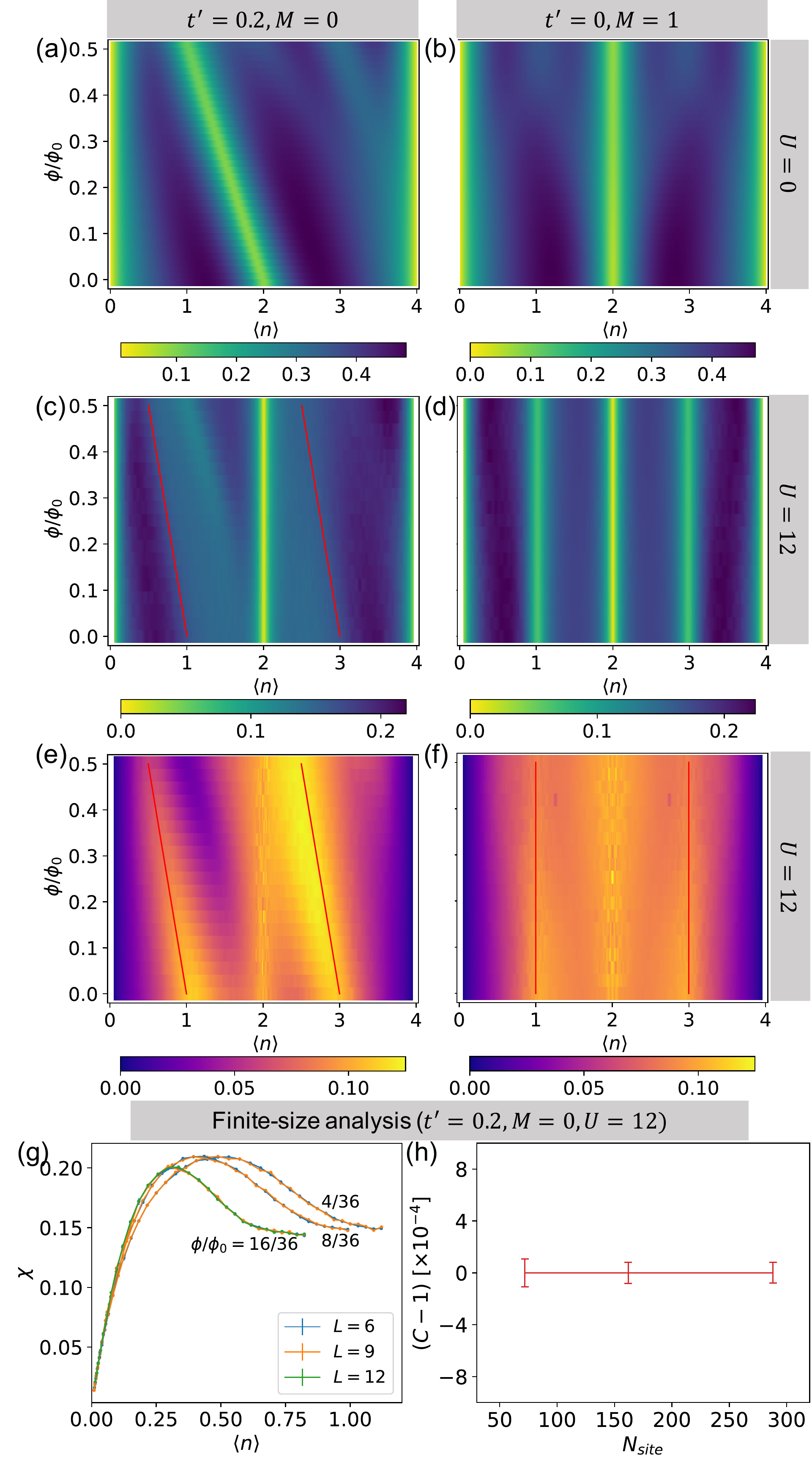}
    \caption{Compressibility $\chi$ as a function of magnetic flux and electron density in a $N_{\text{site}}=6\times6\times2$ cluster for panel (a-d). The parameters are (a,c,e) $t'/t= 0.2,M=0$ and (b,d,f) $t'/t= 0,M=1$. The first and second rows are for the non-interacting and interacting ($U/t=12$) systems, respectively. The third row shows the spin correlation at different interaction strengths $U$. Panel (g) shows the compressibility at low density for various cluster sizes $N_\text{site}=L\times L\times2$ under different magnetic fluxes as labeled. Panel (h) presents the Chern number extracted at different cluster sizes with error bars. The temperature for all cases is $\beta=3/t$.
    }
    \label{fig:HHH1}
\end{figure}

To analyze the phases that emerge, we study the charge compressibility
\beq
\chi=\beta\chi_c=\frac{\beta}{N}\sum_{{\bf i},{\bf j}}\left[ \langle n_{\bf i} n_{\bf j}\rangle - \langle n_{\bf i}\rangle \langle n_{\bf j}\rangle \right],
\eeq
in the presence of Hubbard interactions, where $\chi_c$ is the charge correlation function as a function of external flux $\phi/\phi_0$. The green features in \figdisp{fig:HHH1} occur at integers satisfying the Diophantine equation
\beq
\langle n\rangle=r \frac{\phi}{\phi_0}+s,
\eeq
in which $\langle n\rangle=\langle N_\text{e}\rangle/N_c$, $\langle N_\text{e}\rangle$ means the total number of electrons, $r$ is an integer given by the inverse slope of the straight lines and $s$ is the offset given by the intercept. $r$ defines the Chern number.  Limited by the sign problem (see supplement), we work at relatively high temperature of $\beta=3/t$.
Panels \figdisp{fig:HHH1}(a) and (b) show the results for the compressibility of the non-interacting system. The green lines represent the leading minima in the charge compressibility and hence the intercept at vanishing flux represents the incompressible states at zero field. Note if we were to lower the temperature,  a panoply of quantum Hall states would emerge which are not relevant to the study here as we are focused on the true zero-field features.  As expected in the non-interacting case, the slope vanishes unless the complexified hopping $t'\ne 0$.  In both cases, the green line interpolates to an incompressible state at half-filling, $\langle n\rangle=2$.  For $t'=0.2, M=0$, once strong interactions are turned on, new insulating states indicated by the valleys labeled by red lines in \figdisp{fig:HHH1} (c) appear.  They connect to $\langle n\rangle=1$ and $\langle n\rangle=3$, that is, $1/4$- and $3/4$-filling at zero field. The slope of the red lines gives a Chern number $C=-1$, while at $1/2$-filling, the vertical green line gives $C=0$ for a topologically trivial Mott insulator. For $t'=0, M=1$, strong correlation also introduce new insulating states at $1/4$- and $3/4$-filling, but all the insulating states have vanishing Chern numbers as evidenced by the three vertical lines in \figdisp{fig:HHH1} (d).

We now show that the dips in the charge compressibility are non-trivial because they are correlated
 with peaks in the spin correlation:
\beq
\chi_s=\sum_{m,n,\bf r}S_{m,n}({\bf r})=\frac{1}{N}\sum_{m,n,\bf r}\sum_{{\bf i}_m,{\bf i}_n}\langle S^z_{{\bf i}_m+{\bf r}} S^z_{{\bf i}_n}\rangle,\label{spin}
\eeq
where $m,n=$A,B and ${\bf r}$ represents the vector delineating the unit cell, shown in \figdisp{fig:HHH1}(e) and (f). At $1/4$- and $3/4$-filling, we see distinct peaks in the spin correlation indicating a possible tendency for ferromagnetic alignment.  From panel \figdisp{fig:HHH1}(c), we see that the slope of ridge is also $-1$ from $1/4$- and $3/4$-filling. It indicates that the spin correlation has a peak  exactly at the same place where the compressibility has a valley, as a distinct signature of topological Mott insulator.   This cannot be an accident as it persists for all parameters studied as long as $t'\ne 0$.  When $t'=0$, vertical lines obtain signalling a vanishing of the Chern number.  Strictly speaking, in \figdisp{fig:HHH1} (c) and (e) for a large $U$, what we observe is the precursor of the quarter-filled incompressible states at high temperature. Slightly lower temperature can be reached only at low density and high field (see supplement), showing a more explicit dip to support the red line in \figdisp{fig:HHH1} (c). Indeed as restricted by the Fermion sign problem, we would not be able to access the ground state directly at zero temperature to measure the bulk insulating gap. However, from \figdisp{fig:HHH1} (a) to (c), we have demonstrated that the non-interacting topological edge states at half-filling evolve into the topological Mott edge states at quarter-fillings when $U$ becomes large. In \figdisp{fig:HHH1} (g), the compressibility across different system sizes collapse at least in the density region covering the leading quantum Hall effect (red line) in panel (c). The collapsed curve exhibits negligible finite-size effects, leading to the same Chern number across different system sizes, shown in panel (h), with the error bar from the respective compressibility measurement. Consequently, our simulations are representative of the thermodynamic limit.  The corroboration of the Chern index in the HK model is significant because {\it a priori}, there is no reason to expect an interacting model in which momenta are mixed (the Haldane-Hubbard model) to have the same Chern numbers when the mixing is absent as in the Haldane-HK model.  This tells us that the Chern number is independent of the momentum mixing and as a consequence, the Haldane-HK Hamiltonian \disp{HHK} accurately describes the ground state of the strongly interacting system.  This provides clear evidence that the $1/4-$filled interacting Haldane model is a topological Mott insulator as demonstrated in \figdisp{fig:QHEphase}(d). This is the major point of this paper.

\begin{figure}[ht]
    \centering
    \includegraphics[width=0.46\textwidth]{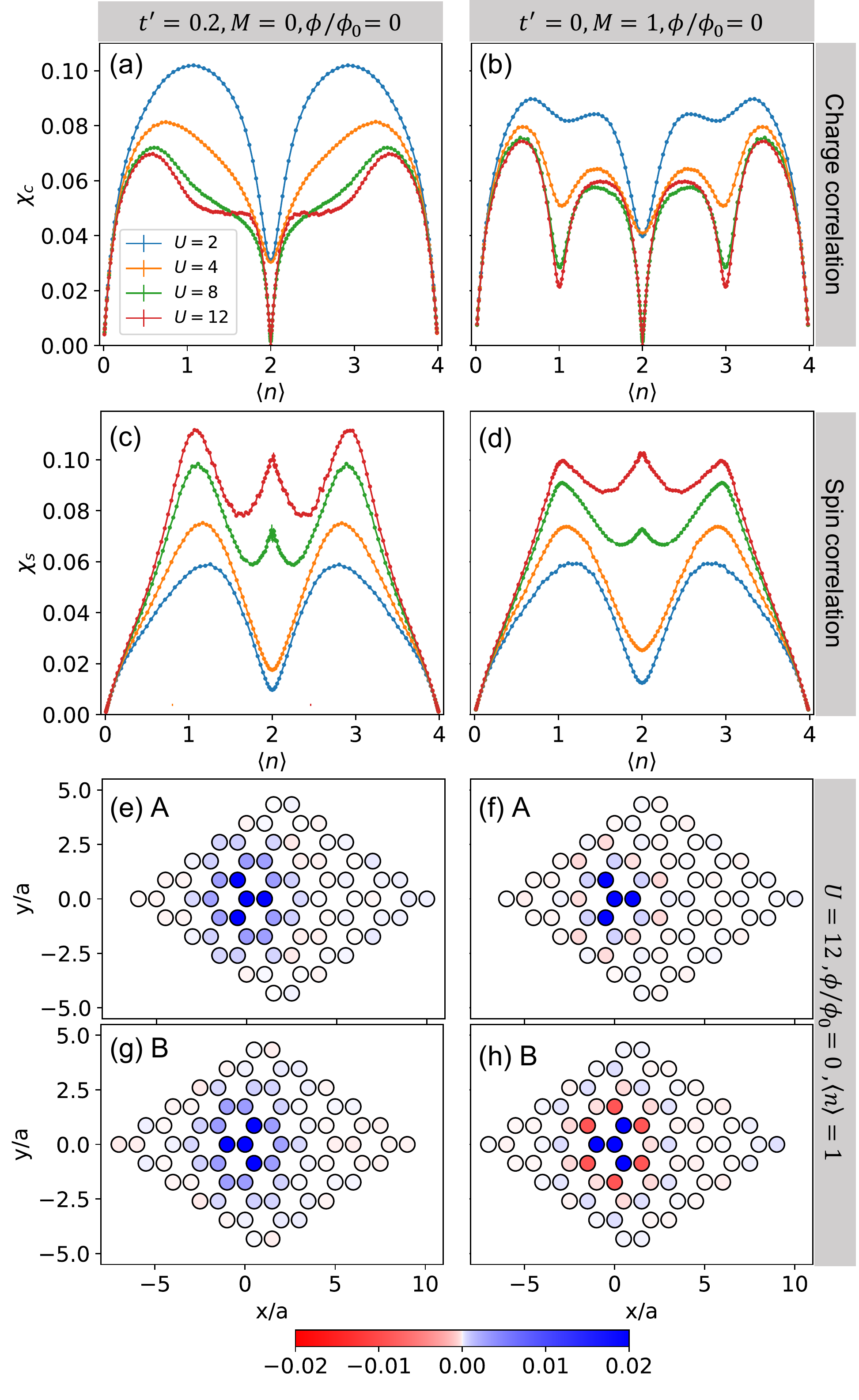}
    \caption{Charge (a,b) and Spin (c,d) correlation at zero external magnetic flux for varying interaction strength, $U$ in a $N_{\text{site}}=6\times6\times2$ cluster. Panels (a-d) share the same legend. The parameters are (a,c,e,g) $t'/t= 0.2,M=0$ and (b,d,f,h) $t'/t= 0,M/t=1$. The third and fourth rows show the real-space zero-frequency spin susceptibility pattern at $1/4$-filling $\langle n \rangle=1$ from the perspective of the A and B sublattices, respectively. The temperature is $\beta=3/t$ for all panels.
    }
    \label{fig:HHH2}
\end{figure}

We next focus on the physics at the zero-field limit. \figdisp{fig:HHH2} (a) and (b) represent the charge correlation $\chi_c=\chi/\beta$ as a slice through \figdisp{fig:HHH1} (c) and (d) respectively at zero flux, $\phi/\phi_0=0$.  We see clearly that as the Hubbard interaction increases, the features outlined by the red line in \figdisp{fig:HHH1}(c) turn into distinct dips for both $t'\ne 0$ and for the topologically trivial case, $t'=0$.  This trend is more sharply apparent in \figdisp{fig:HHH2} (b), however. This state of affairs obtains because the Hubbard interaction mixes the Haldane basis for $t'\ne 0, M=0$ but not the sublattice states generated from $t'=0, M \ne 0$.  Nonetheless, in both cases the dips in the charge compressibility, which decrease as the temperature is lowered, are consistent with incompressible states at  $1/4$- and $3/4$-filling.  These dips in the charge compressibility are correspondent with peaks in the spin correlation as demonstated in
\figdisp{fig:HHH2}(c) and (d). In both panels (c) and (d), as $U$ increases, a peak in spin correlation  develops at $1/4$-(or $3/4$-)filling where a dip appears in the compressibility (charge correlation).  As remarked earlier this cannot be an accident and lends further credence to the existence of a phase transition at $1/4$-(or $3/4$-)filling. Note although a peak also appears at half filling for large $U=8,12$ in both panels, it turns into a dip at lower temperature since the half-filled system becomes antiferromagnetic (see supplement).  Finally, to detail the spatial dependence of the spin correlation at $1/4$-filling, we calculate the real-space zero-frequency spin susceptibility
\beq
\begin{aligned}
S_{m,n}({\bf r},\omega=0)=\frac{1}{N}\sum_{{\bf i}_m,{\bf i}_n}\int_0^{\beta}\langle S^z_{{\bf i}_m+{\bf r}}(\tau) S^z_{{\bf i}_n}(0)\rangle d\tau,\label{spin2}
\end{aligned}
\eeq
which is more sensitive to short-range fluctuating order at high temperature than the equal-time spin correlation in \disp{spin}.  We plot $S_{m,A}({\bf r},\omega=0)$ and $S_{m,B}({\bf r},\omega=0)$ in panel (e,g) respectively for $t'\ne 0, M=0$ and panel (f,h) respectively for $t'=0, M\ne0$. The topologically non-trivial case, \figdisp{fig:HHH2} (e,g), reveals ferromagnetic correlations on both sublattices. Note that the ferromagnetic insulating state at $1/4$-filling is different from the quantum Hall ferromagnetic states in \refdisp{Mai} although they both come with an odd-integer Chern number and a peak on spin correlation at ${\bf q}=0$. Here the ferromagnetic insulator appears at zero field and the correlation decays with increasing flux, shown in \figdisp{fig:HHH2}(a), while the quantum Hall ferromagnetic states in \refdisp{Mai} requires finite magnetic flux and vanishes as the flux decreases. In the topologically trivial case, we first note a distinct difference between panels \figdisp{fig:HHH2}(f) and (h), implying a lifting of the sublattice degeneracy. While panel (f) shows ferromagnetism, panel (h) exhibits something more akin to a spin-density wave. 

\begin{figure}[ht]
    \centering
    \includegraphics[width=0.5\textwidth]{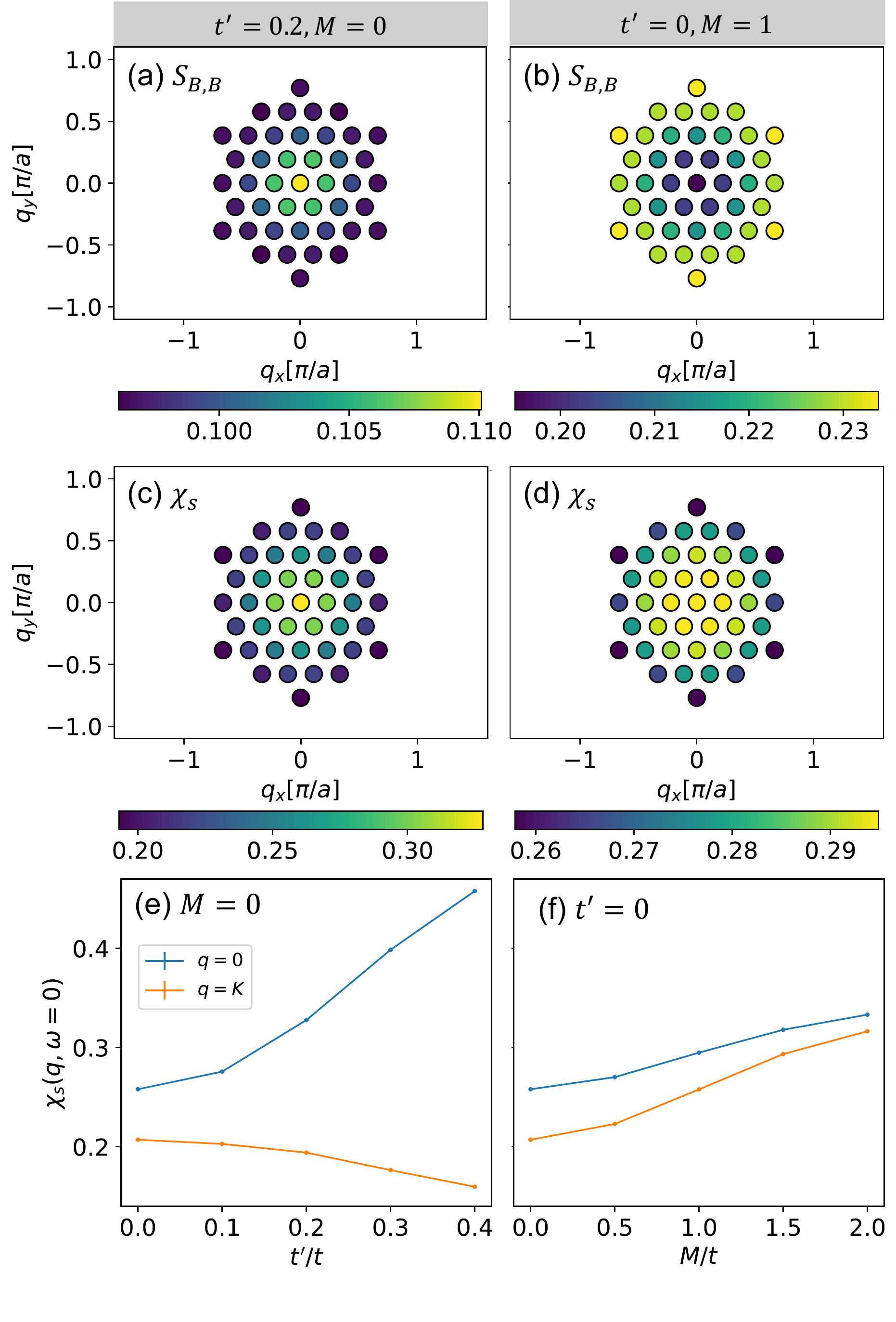}
    \caption{Zero-frequency spin susceptibility at quarter-filling. The first and second rows show the spin susceptibility pattern in momentum space $S_{B,B}({\bf q},\omega=0)$ within the B sublattice and $\chi_s({\bf q},\omega=0)$ summing over all sublattices for different parameter sets: (a) $t'/t= 0.2,M=0$ and (b) $t'/t= 0,M=1$. In the third row, panels (e) and (f) contain the zero-frequency spin susceptibility at ${\bf q}=0$ and ${\bf q}=K$ as a function of $t'$ (fixed $M=0$) and $M$ (fixed $t'=0$), respectively. The other parameters are $U/t=12$ and $\beta=3/t$ for all panels. }
    \label{fig:HHH3}
\end{figure}

To investigate this possibility further, we focus on the correlations in momentum space defined as 
\beq
\begin{aligned}
&\chi_s({\bf q},\omega=0)
\\&=\sum_{m,n} S_{m,n}({\bf q},\omega=0)
\\&=\sum_{m,n}\sum_{\bf r}S_{m,n}({\bf r},\omega=0)\exp[-i {\bf q}\cdot({\bf r}+\Delta{\bf r}_{m,n})].
\end{aligned}
\eeq
This is displayed in \figdisp{fig:HHH3}. Specifically, we present $S_{B,B}({\bf q},\omega=0)$ and $\chi_s({\bf q},\omega=0)$ in the first and second rows, respectively. Panels \figdisp{fig:HHH3} (a) and (c) show that regardless of the sublattice, the spin correlations are peaked at ${\bf q}=0$ when $t'\ne 0, M=0$. In \figdisp{fig:HHH3} (e) as $t'$ increases, the ferromagnetic (${\bf q}=0$) spin correlation is enhanced, while the correlation at the K point $K=(0,4\pi/3\sqrt{3}a)$ dies out. Consequently, only ${\bf q}=0$ ferromagnetism survives in the ground state of the $1/4$-filled state. This phenomenon is essentially the same for $3/4$-filled state and as a consequence not shown. Alternatively, in the topologically trivial case, we find that within sublattice B, \figdisp{fig:HHH3} (b) indicates that the spin susceptibility is peaked at the $\bf K$ point and the symmetry related points in the Brillouin zone, while the other components $S_{A,A}$, $S_{A,B}$ and $S_{B,B}$ have a peak at ${\bf q}=0$ (not presented). After summing over all the components, we see that $\chi_s({\bf q},\omega=0$ in \figdisp{fig:HHH3} (d)  reveals that the contributions at the $K$ point and the center of the Brillouin zone are essentially equal at this temperature, given the short range in the color bar. \figdisp{fig:HHH3} (f) shows that as $M$ increases, the difference of spin correlation between the origin and K point becomes smaller, thereby showing no tendency for any particular order.

\section{Experimental Realizations}

Ultra-cold atom systems provide a natural venue to search for topological Mott physics due to the high degree of control over microscopic parameters they afford.  In particular, such systems have been used to realize the non-interacting Haldane Hamiltonian including with spin mixtures\cite{haldaneexp}. Given the flexibility of this platform to effect fermionic interactions with tunable strength\cite{hulet,EsslingerReview}, a realization of the interacting Haldane model is in principle feasible. In addition, single layers of AFe$_2$(PO$_4$)$_2$ (A=Ba,Cs,K,La) have been shown\cite{kee} to be described by the Haldane model. However, because the spin bands are non-degenerate at the single-particle level, a pseudospin degree of freedom would be required to realize topological Mott physics, such as that offered by the three single-layer planes in the bulk unit cell. Finally we note that several moir\'e systems have been predicted\cite{WuHomobilayer,flat} and observed\cite{Mak} to exhibit quantum anomalous Hall phases that map onto the spinless Haldane model. Stacking of two strongly interacting but electrically isolated copies of these superlattices, e.g. by twist decoupling or using a hexagonal boron nitride spacer whose thickness is small compared to the screening length, could also generate an effective layer pseudospin capable of supporting Mott physics in the topological bands.

\section{Physical Interpretation: $4=1+1+1+1\neq 2+2$}

While there have been studies of the 1/4-filled Haldane-Hubbard model previously\cite{qf1,qf2}, none have disclosed both the topology and Mottness of that insulator.  That quarter filling is ideally suited for Mottness follows from the 2-band nature of the spin-degenerate bands in the Haldane model.  Consider a non-topological system containing 2 atoms per unit cell. The ground state is now a band insulator in which the lower band is filled.  If interactions are now added to this system,  a Mott insulator ensues in the lower band at 1/4-filling of the lower band.  It is for this reason that it matters not the form of the interactions,  local-in momentum space or local-in real space, neither of which destroys the underlying topology of the Haldane bands.   All that matters is that $4=1+1+1+1$.  In a standard Mott insulator, the spectral weight at each k-point is split over two bands, the upper and lower Hubbard bands.  In the topological Haldane model, each $k-$state is split over two bands.  In the absence of interactions, each $k-$state can be doubly occupied with no energy cost, giving rise to $4=2+2$ When interactions are included, the singly and doubly occupied bands now have distinct energies as indicated by the 4 poles in the Green function, hence the equation $4=1+1+1+1\ne 2+2$. The standard Mott picture consists just of two bands.  Now, however, 4 bands exist each with equal spectral weight.  The two lowest singly-occupied bands (unmeshed blue and orange in \figdisp{fig:QHEphase}) remain singly occupied with a fixed energy separation of the topological gap $\Delta$.  As a result, the 1/4-filled state is gapped with non-trivial topological signature, $C=-1$.  It is for this reason that the Hubbard-Haldane model yields the same results because Mott insulation in the presence of topology (at least that in the Haldane model) necessarily gives rise to $4$ non-degenerate energy bands.  Since these bands are inherently in momentum space, 
the Mott insulation is more easily understood in momentum space rather than in real space.  At work here, as we have demonstrated\cite{HKnp2} all interactions break the local-in momentum space $Z_2$ symmetry of a Fermi liquid.  The HK interaction is the simplest one that does this and hence suffices to explain the origin of topological Mott insulation. Undoubtedly, extensions of this work beyond the Haldane model exist.  Nonetheless, this work provides a general framework for thinking about this new state of matter.

\section{Acknowledgements}

Supported by the Center for Quantum Sensing and Quantum Materials, a DOE Energy Frontier Research Center, under award DE-SC0021238 (P.M. B. E. F., and P.W.P.)  PWP also acknowledges NSF DMR-2111379 for partial funding of the HK work which led to these results and Srinivas Raghu for comments on an earlier draft. The DQMC calculation of this work used the Advanced Cyberinfrastructure Coordination Ecosystem: Services \& Support (ACCESS) Expanse supercomputer through the research allocation TG-PHY220042, which is supported by National Science Foundation grant number ACI-1548562\cite{xsede}.

\appendix

\section{Hamiltonian with Interaction-Independent Basis}
The analytically solvable Hamiltonian in Eq.~(1) of the main text provides a simple model for the interacting Haldane electrons. However, the interaction term has a simple form (leading to the simple solution) only in the basis that diagonalizes the non-interacting Hamiltonian.  This means that the interaction term depends on the choice of $t'$, $\psi$ and $M$. The natural question is whether or not it is possible to write the interaction term in a similarly simple form independent of basis without surrendering the essential solution outlined in the text. The answer is yes. We can write the  Hamiltonian as 
\begin{equation}
\begin{split}
H&=\sum_{{\bf k},\sigma}\big[(\varepsilon_{+,{\bf k}}-\mu)n_{+,{\bf k}\sigma}+(\varepsilon_{-,{\bf k}}-\mu)n_{-,{\bf k}\sigma}\big]
\\&+U^\prime\sum_{\bf k}(n_{+,{\bf k}\uparrow}+n_{-,{\bf k}\uparrow})(n_{+,{\bf k}\downarrow}+n_{-,{\bf k}\downarrow}). \label{HHK2}
\end{split}
\end{equation}
The quantity $n_{+,{\bf k}\sigma}+n_{-,{\bf k}\sigma}$ is a trace in the Haldane orbital space and is therefore independent of the orbital basis, that is, the interaction term is the same for all combinations of $t'$, $\psi$ and $M$. This Hamiltonian is also analytically solvable, since the interaction does not mix the original non-interacting eigenstates. We can write the Green function as

\begin{equation*}
\begin{aligned}
&G_{\pm,{\bf k}\sigma}(i\omega_n\rightarrow \omega)\\&=\frac{\langle(1- n_{+,{\bf k}\bar{\sigma}})(1- n_{-,{\bf k}\bar{\sigma}})\rangle}{\omega-\varepsilon_{\pm,{\bf k}}} 
\\&+ \frac{\langle(1- n_{+,{\bf k}\bar{\sigma}}) n_{-,{\bf k}\bar{\sigma}}\rangle+\langle(1- n_{-,{\bf k}\bar{\sigma}}) n_{+,{\bf k}\bar{\sigma}}\rangle}{\omega-(\varepsilon_{\pm,{\bf k}}+U^\prime)}
\\&+\frac{\langle n_{+,{\bf k}\bar{\sigma}} n_{-,{\bf k}\bar{\sigma}}\rangle}{\omega-(\varepsilon_{\pm,{\bf k}}+2U^\prime)}
\\&=\frac{\langle(1- n_{+,{\bf k}\bar{\sigma}})(1- n_{-,{\bf k}\bar{\sigma}})\rangle}{\omega-\varepsilon_{\pm,{\bf k}}} 
\\&+\frac{\langle n_{+,{\bf k}\bar{\sigma}}\rangle
+\langle n_{-,{\bf k}\bar{\sigma}}\rangle-2\langle n_{+,{\bf k}\bar{\sigma}} n_{-,{\bf k}\bar{\sigma}}\rangle}{\omega-(\varepsilon_{\pm,{\bf k}}+U^\prime)}
\\&+\frac{\langle n_{+,{\bf k}\bar{\sigma}} n_{-,{\bf k}\bar{\sigma}}\rangle}{\omega-(\varepsilon_{\pm,{\bf k}}+2U^\prime)}.
\label{GF2}
\end{aligned}
\end{equation*}

\begin{figure}[ht]
    \centering
    \includegraphics[width=0.4\textwidth]{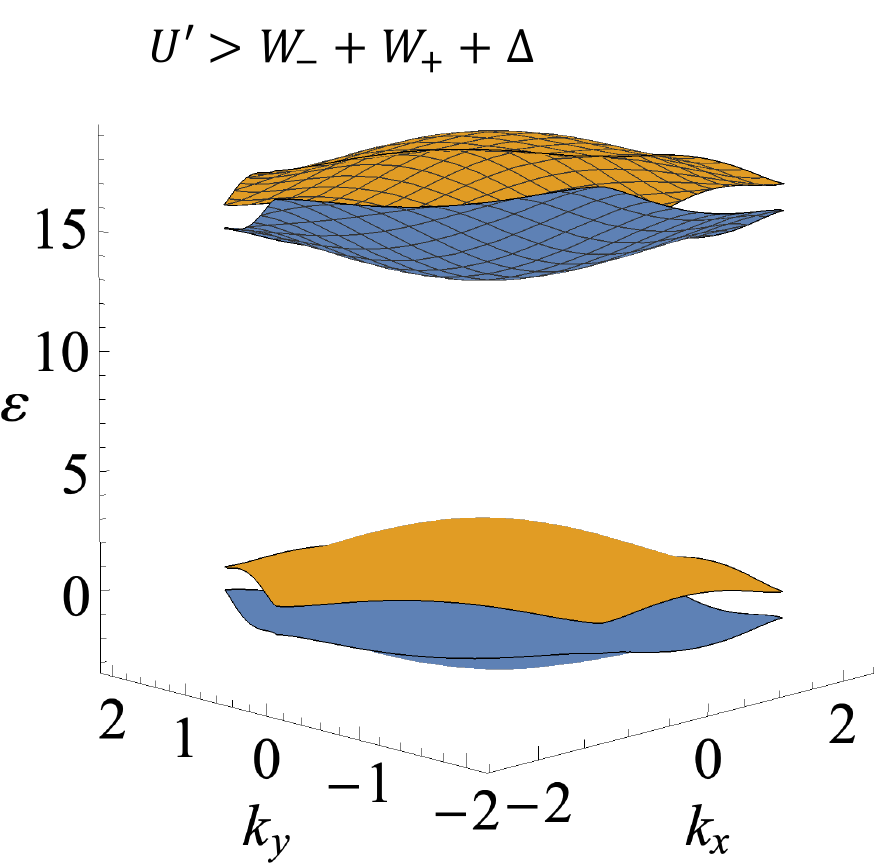}
    \caption{Band structure for the generalized Haldane-HK model in \disp{HHK2} with $t'/t=0.1, \psi=-\pi/2, M=0$ and $U'=8>W_++W_-+\Delta$. $\Delta=6\sqrt{3}t'\sin{\psi}$ is the non-interacting topological gap. $W_{+(-)}$ correspond to the bandwidth of the Haldane upper (lower) band. The blue (or orange)  color represents Chern number $C=-1 ($or $ 1)$. The meshed band consists of only doubly occupied states, while the unmeshed band is singly occupied.}
    \label{fig:sup_replotU16}
\end{figure}

At zero temperature with a large enough $U^\prime$ ($U^\prime>W_++W_-+\Delta$), for each ${\bf k}$, the state with the same spin would be filled up first to avoid the cost of interaction energy. After that, the state with opposite spin would start to be filled with the interaction energy cost of $2U^\prime$. In this sense, the band structure would be similar to that for the Hamiltonian in Eq.~(1) of the main text if $U=2U^\prime$. Choosing the same example parameter set $t'=0.1$, $\psi=-\pi/2$ and $M=0$ ($W_++W_-+\Delta=6$), we plot the band structure in \figdisp{fig:sup_replotU16} (setting $U^\prime=8$). The system is a topological Mott insulator at $1/4$- (or $3/4$-) filling and a topologically trivial Mott insulator at $1/2$-filling. The band structure is almost the same as that in Fig.~2(d) of the main text if $U=2U^\prime=16$. The only difference is that in \figdisp{fig:sup_replotU16} for Hamiltonian \disp{HHK2}, once the lowest band is filled up, the states of the second lowest band are fixed with no degeneracy. For each ${\bf k}$, the state contains the same spin as that in the lowest band with the same momentum. In Fig.~2(d) of the main text, the degeneracy of the second lowest band remains with a full lowest band. In short, we construct a Hamiltonian \disp{HHK2} for interacting Haldane electrons independent of basis and containing essentially the same physics as Eq.~(1) of the main text under strong correlation.

\section{Determinantal quantum Monte-carlo simulation for the Haldane-Hubbard model}

The determinantal quantum Monte-carlo (DQMC) simulation conducted for Haldane-Hofstadter-Hubbard model is similar to that for the single-band Hubbard model\cite{Blankenbecler,Hirsch,White}. The only difference is the use of complex number for physical quantities due to the complex second neighbor hopping and the Peierls phase. 

\begin{figure}[ht]
    \centering
    \includegraphics[width=0.5\textwidth]{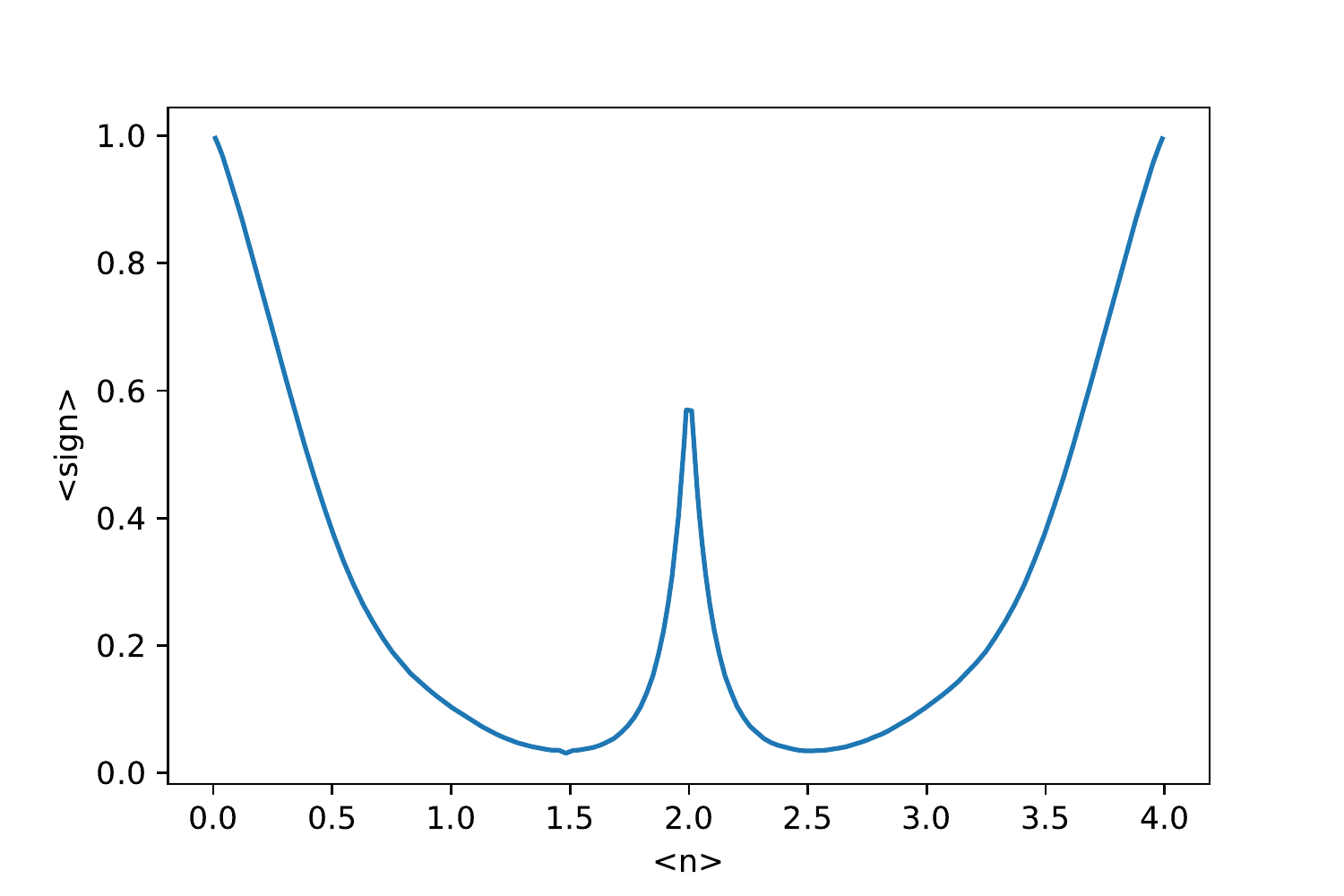}
    \caption{Average DQMC sign as a function of density for $t'=0.2, M=0, U=12$ and $\beta=3$ at zero magnetic flux.}
    \label{fig:sign}
\end{figure}

DQMC suffers from a fermionic sign problem when turning on the Hubbard interaction in the Haldane-Hubbard model. The Hubbard-Stratonovich transformation is not SU(2) symmetric so that the half-filling has a better sign problem. Generally, the average sign decays exponentially as the interaction strength increases or temperature decreases, leading to the exponential increase on the necessary number of measurements to bring down the error bar. For this reason, we are restricted at a relatively high temperature $\beta=3$ while working with a large $U=12$. The average sign varies largely along with the density, as shown in \figdisp{fig:sign}.

We discretize the imaginary time $\beta$ into $L$ slides with $\Delta\tau=0.1$. We conduct 10000 warmup sweeps and 200000 measurement sweeps (10 measurements per sweep) at each Markov chain. We scan the chemical potential from $-10$ to $10$ (about 160 values) to obtain the density dependence of physical quantities. Depending on the sign problem in \figdisp{fig:sign} (little change under finite magnetic field), we use different numbers of Markov chain to bring down the error bar. 40 Markov chains are for the worst case around $\langle n \rangle=1.5, 2.5$. We use equal-time and unequal-time measurement to produce Fig.~3 and 4(a-d), and unequal-time measurement for Fig.~4(e-h) and ~5, of the main text.

\section{Lower temperature}
\begin{figure}[ht]
    \centering
    \includegraphics[width=0.5\textwidth]{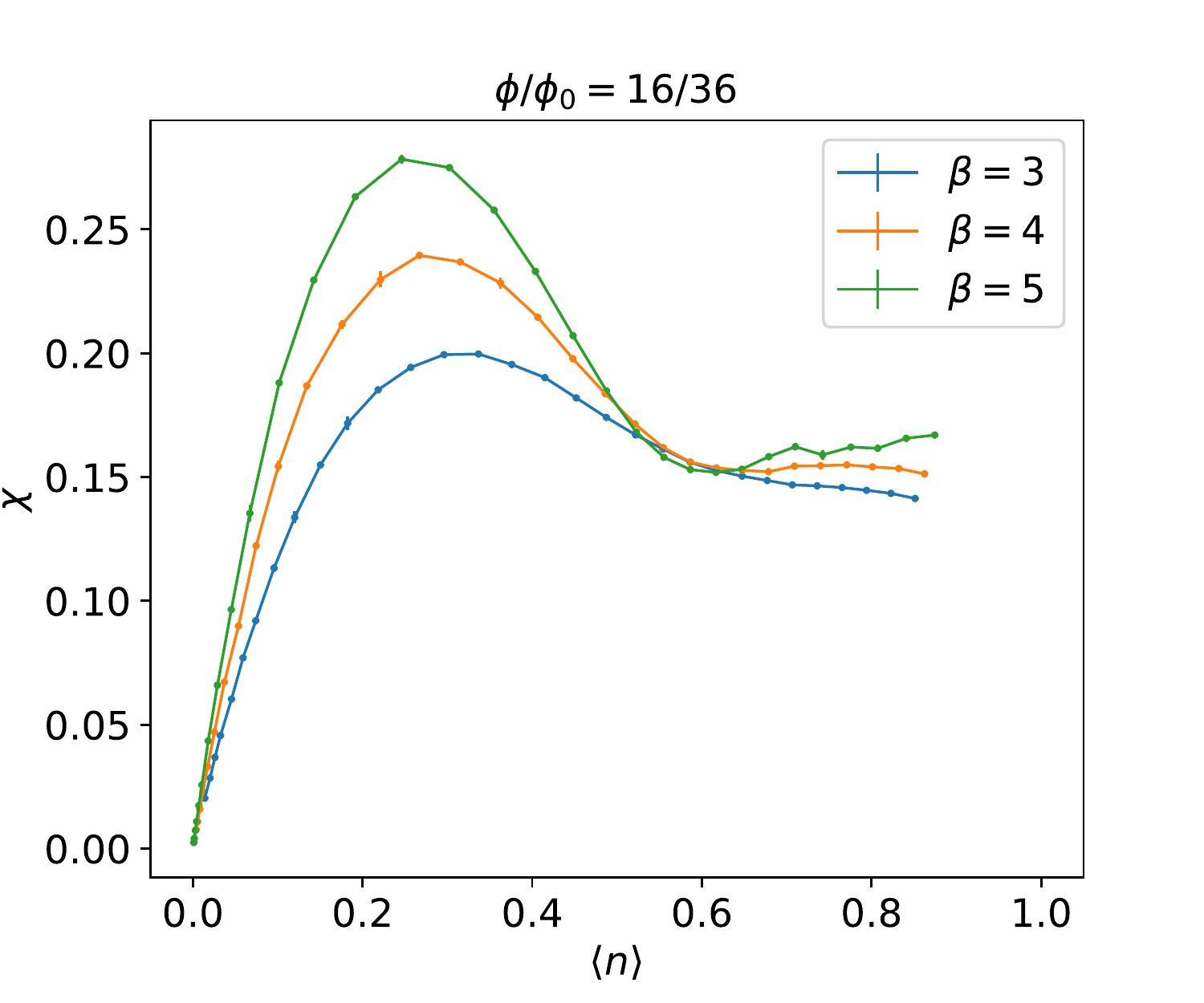}
    \caption{The compressibility as a function of density for $\phi/\phi_0=16/36$ at different inverse temperatures $\beta=3,4,5$.}
    \label{fig:lowerT}
\end{figure}

The sign problem shown in \figdisp{fig:sign} prevents us from getting reliable results at lower temperature for all densities. However, under high field where the density is low but enough to cover the interested region, we can carry out the calculation to a lower temperature, as shown in \figdisp{fig:lowerT}. As temperature decreases, the dip representing the anomalous quantum Hall effect from $1/4$-filling becomes clearer. 

\section{Anti-ferromagnetism in the half-filled system}
\begin{figure}[ht]
    \centering
    \includegraphics[width=0.5\textwidth]{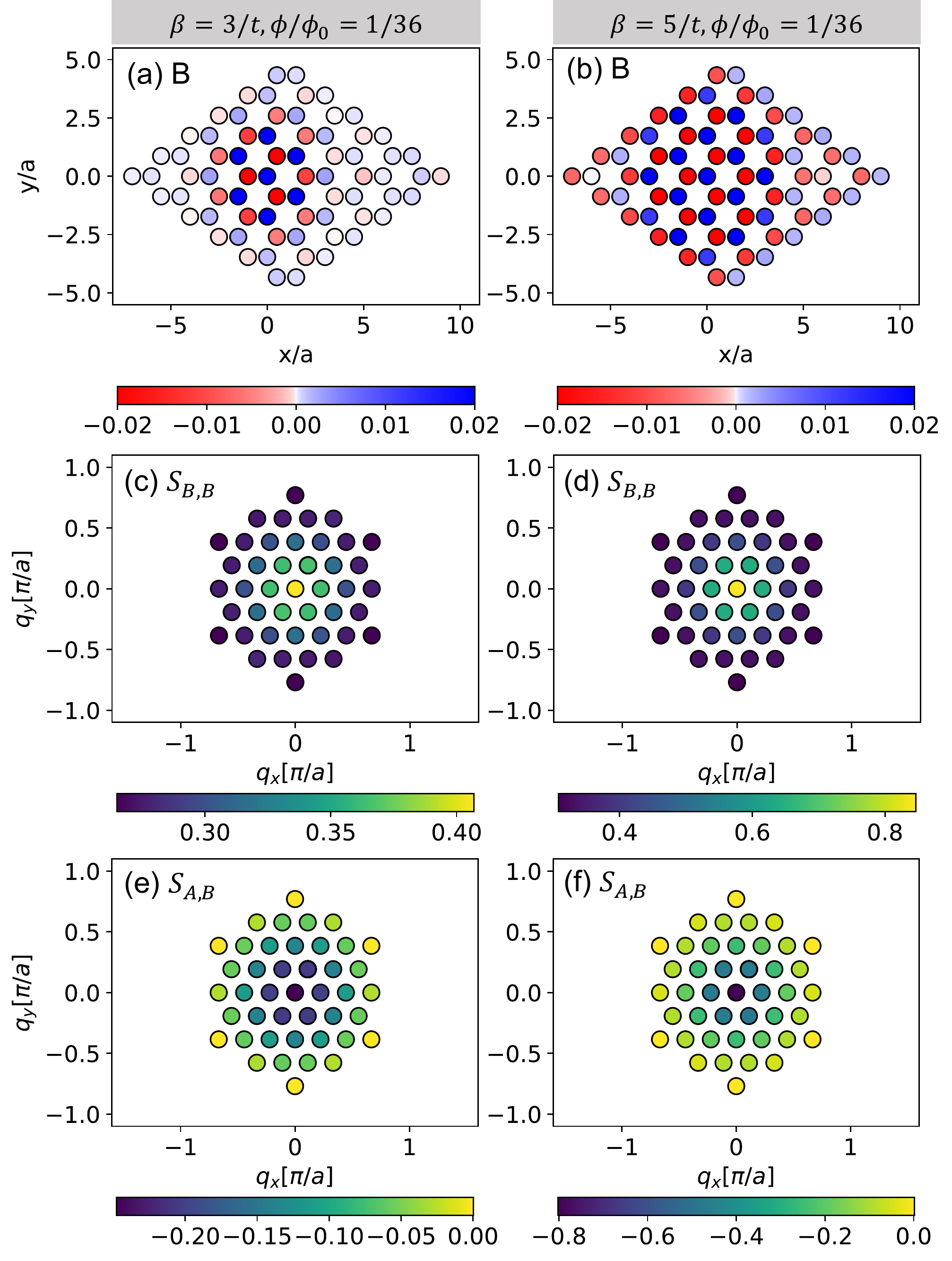}
    \caption{Zero-frequency spin susceptibility at half-filling for the topologically nontrivial case $t'=0.2, M=0$. The left and right columns show the susceptibility at the temperature $\beta=3/t$ and $\beta=5/t$, respectively. The first row features $S_{m,B}({\bf r},\omega=0)$ ($m=$A,B) in real space. The second and third rows display $S_{B,B}({\bf q},\omega=0)$ and $S_{A,B}({\bf q},\omega=0)$ in momentum space, respectively. The interaction strength is $U/t=12$ and the smallest magnetic flux $\phi/\phi_0=1/36$ is applied for all panels.}
    \label{fig:tp02M0}
\end{figure}

\begin{figure}[ht]
    \centering
    \includegraphics[width=0.5\textwidth]{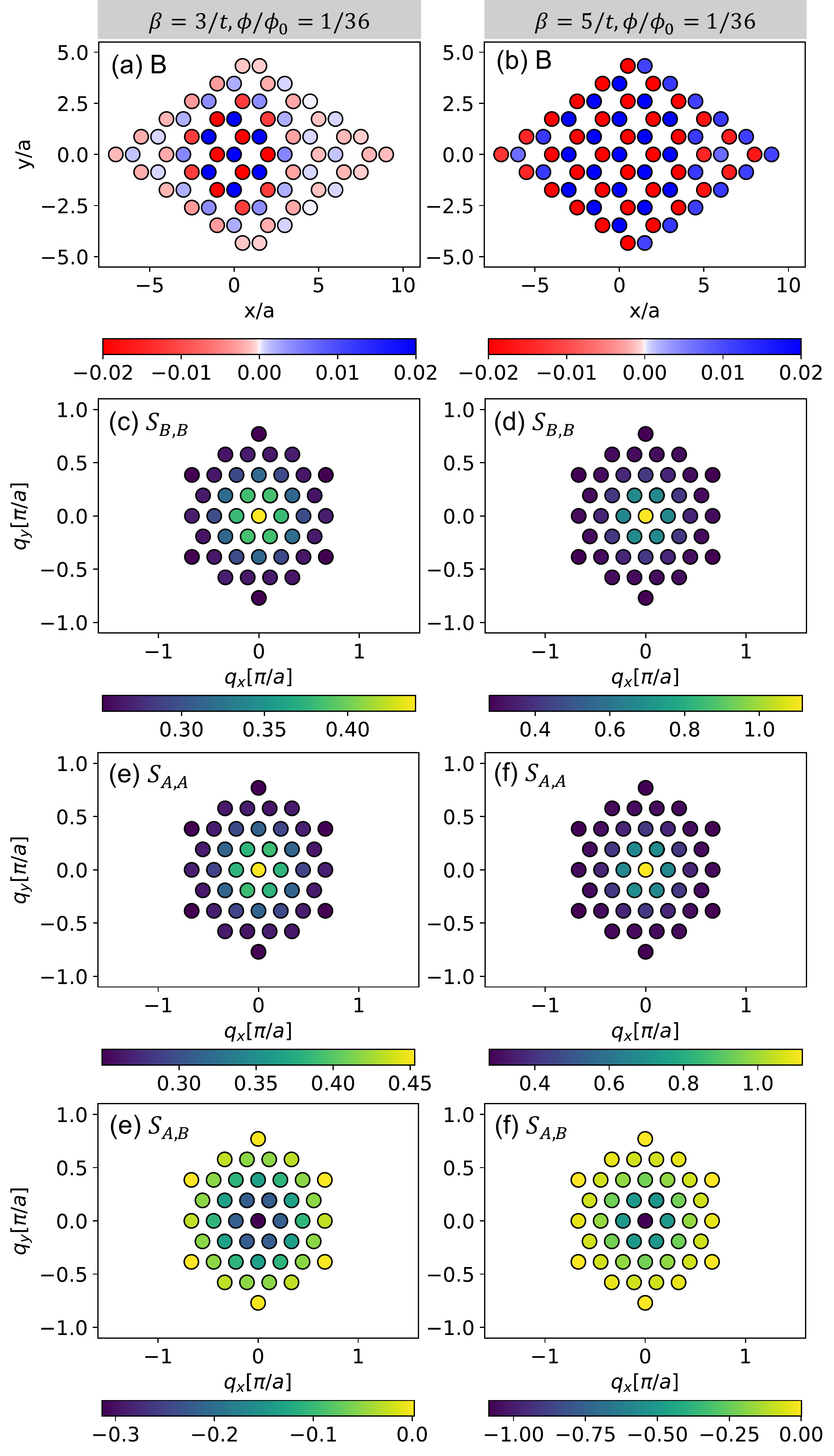}
    \caption{Zero-frequency spin susceptibility at half-filling for the topologically trivial case $t'=0, M=1$. The left and right columns show the susceptibility at the temperature $\beta=3/t$ and $\beta=5/t$ respectively. The first row displays $S_{m,B}({\bf r},\omega=0)$ ($m=$A,B) in real space. The second, third rows show $S_{B,B}({\bf q},\omega=0)$, $S_{A,A}({\bf q},\omega=0)$ and $S_{A,B}({\bf q},\omega=0)$ in momentum space, respectively. The interaction strength is $U/t=12$ and the smallest magnetic flux of $\phi/\phi_0=1/36$ is applied for all panels.}
    \label{fig:tp0M1}
\end{figure}

To explore the structure of spin correlation at  half filling, the middle peak of Fig.~4 (c) and (d) in the main text, we calculate the zero-frequency spin susceptibility in real and momentum space:
\beq
\begin{split}
S_{m,n}({\bf r},\omega=0)=\frac{1}{N}\sum_{{\bf i}_m,{\bf i}_n}\int_0^{\beta}\langle S^z_{{\bf i}_m+{\bf r}}(\tau) S^z_{{\bf i}_n}(0)\rangle d\tau,\label{rspace}
\end{split}
\eeq
\beq
\begin{split}
&S_{m,n}({\bf q},\omega=0)\\&=\sum_{m,n}\sum_{\bf r}S_{m,n}({\bf r},\omega=0)\exp[-i {\bf q}\cdot({\bf r}+\Delta{\bf r}_{m,n})].\label{kspace}.
\end{split}
\eeq
The results  are shown in \figdisp{fig:tp02M0} for the topologically nontrivial case ($t'=0.2,M=0$) and \figdisp{fig:tp0M1} for the topologically trivial case ($t'=0,M=1$) at $\beta=3/t$ and $\beta=5/t$. Note that the smallest magnetic flux $\phi/\phi_0=1/36$ is applied to reduce the finite size effects at lower temperature, and hence the conclusion can be extrapolated to zero field. In \figdisp{fig:tp02M0}, the first row features $S_{m,B}({\bf r},\omega=0)$ ($m=A,B$) indicating that anti-ferromagnetism develops as the temperature decreases. For  strong anti-ferromagnetism, the spin correlation should be ferromagnetic in the intra-sublattice components and anti-ferromagnetic in the inter-sublattice components with the same amplitude. Therefore, we show $S_{B,B}({\bf q},\omega=0)$ and $S_{A,B}({\bf q},\omega=0)$ in the second and third rows, respectively. At $\beta=3/t$, indeed $S_{B,B}({\bf q},\omega=0)$ is ferromagnetic and $S_{A,B}({\bf q},\omega=0)$ is anti-ferromagnetic. But the amplitude yields $|S_{A,B}({\bf q}=0,\omega=0)|<|S_{B,B}({\bf q}=0,\omega=0)|$ leading to a residual ${\bf q}=0$ contribution in the total spin correlation corresponding to the middle peak in Fig.~3(c) in the main text. However, at a lower temperature $\beta=5/t$, the amplitude becomes comparable $|S_{A,B}({\bf q}=0,\omega=0)|\approx|S_{B,B}({\bf q}=0,\omega=0)|$, resulting in a vanishing ${\bf q}=0$ component for the total spin correction, as expected for anti-ferromagnetism. The situation is similar for the topologically trivial case in \figdisp{fig:tp0M1} except that now we are comparing the amplitude $|S_{A,B}({\bf q}=0,\omega=0)+S_{B,A}({\bf q}=0,\omega=0)|$ and $|S_{A,A}({\bf q}=0,\omega=0)+S_{B,B}({\bf q}=0,\omega=0)|$ due to the lifting of the sublattice degeneracy. Similarly, $S_{A,B}({\bf q}=0,\omega=0)$=$S_{B,A}({\bf q}=0,\omega=0)$ for symmetry reasons.

\bibliography{reference}




\end{document}